\def\ARXIVCOPY{1}
\documentclass{article}

\usepackage[T1]{fontenc}
\IfFileExists{iclr2027_conference.sty}{%
  \usepackage{iclr2027_conference,times}%
}{%
  \usepackage[margin=1in]{geometry}%
}
\usepackage{hyperref}
\hypersetup{hidelinks}
\usepackage{graphicx}
\usepackage{float}
\IfFileExists{xurl.sty}{\usepackage{xurl}}{\usepackage{url}}
\usepackage{booktabs}
\usepackage{amsmath}
\usepackage{amssymb}
\usepackage{tabularx}
\usepackage{array}
\usepackage{colortbl}
\usepackage{microtype}
\usepackage{needspace}

\ifdefined\ARXIVCOPY
  \iclrfinalcopy
\fi

\newcolumntype{Y}{>{\raggedright\arraybackslash}X}
\newcolumntype{C}[1]{>{\centering\arraybackslash}p{#1}}
\newcolumntype{L}[1]{>{\raggedright\arraybackslash}p{#1}}
\definecolor{selectedrow}{RGB}{230,240,255}

\title{Text Scores Do Not Establish Performance on Lexically Non-Diagnostic Speech Tasks: A Qwen2-Audio Quantization Case Study}
\ifdefined\ARXIVCOPY
\author{
Mengzhe Geng\thanks{Corresponding author. Email: \texttt{Mengzhe.Geng@nrc-cnrc.gc.ca}.}\\
National Research Council Canada
\And
Jinxi Ji\\
The Hong Kong Polytechnic University
\And
Junhao Xu\\
The Chinese University of Hong Kong
}
\else
\author{Anonymous Authors}
\fi

\begin{document}
\maketitle
\ifdefined\ARXIVCOPY
\fancyhf{}
\renewcommand{\headrulewidth}{0pt}
\else
\fi
\renewcommand{\footrulewidth}{0pt}
\cfoot{\thepage}
\pagestyle{fancy}
\fancypagestyle{plain}{%
  \ifdefined\ARXIVCOPY
    \fancyhf{}
    \renewcommand{\headrulewidth}{0pt}
  \else
    \fancyhf{}
    \fancyhead[L]{Under review as a conference paper at ICLR 2027}
  \fi
  \renewcommand{\footrulewidth}{0pt}
  \cfoot{\thepage}}
\thispagestyle{plain}

\begin{abstract}
Text-output scores alone do not show whether quantization preserves performance on speech tasks whose target labels cannot be recovered from the transcript. We evaluate fixed mixed 4/8-bit Qwen2-Audio-7B-Instruct allocations averaging 6 and 7 bits per parameter on 508 English-to-German FLEURS utterances and on 512 RAVDESS emotion clips from 16 speakers. The BLEU and chrF differences from half precision (FP16) have intervals that include zero for both allocations. On RAVDESS, the same two sentences occur equally often with every emotion label. The absolute accuracy differences from FP16 are $-3.71$\% for 6 bit and $-1.17$\% for 7 bit. The 6-bit speaker interval excludes zero and an exact two-sided sign-flip test gives $p=0.0148$; the 7-bit interval includes zero. Same-budget controls do not identify either selected allocation as best. This case study shows why translation scores and performance on tasks beyond the transcript need separate evaluation.
\end{abstract}

\section{Introduction}

\textbf{A good transcript does not establish performance on an emotion task after quantization.}\ifdefined\ARXIVCOPY\footnote{Public reproducibility repository: \href{https://github.com/MENGZHEGENG/speechllm-quantization}{github.com/MENGZHEGENG/speechllm-quantization}.}\fi\ Two recordings can contain the same words but carry different emotion labels. The transcript therefore cannot determine the label. We call such a task \emph{lexically non-diagnostic}. A model may preserve recognition or translation quality while its emotion accuracy changes. This task property does not identify which acoustic cues the model uses.

Post-training quantization (PTQ) studies commonly report text-task quality alongside memory or speed measurements \citep{dettmers2022llmint8,frantar2022gptq,xiao2022smoothquant,lin2023awq}. Recent speech-model work extends mixed-precision PTQ and compression to speech recognition, speech translation, and speech language models \citep{genptq2025,ultralowptqspeech2025,efficientstcompression2025,dietkit2026,selectedcodec2026}. Word error rate measures transcript accuracy, and translation scores measure text output. Neither evaluates a target label that cannot be recovered from the transcript.

\begin{figure}[t]
\centering
\includegraphics[width=\linewidth]{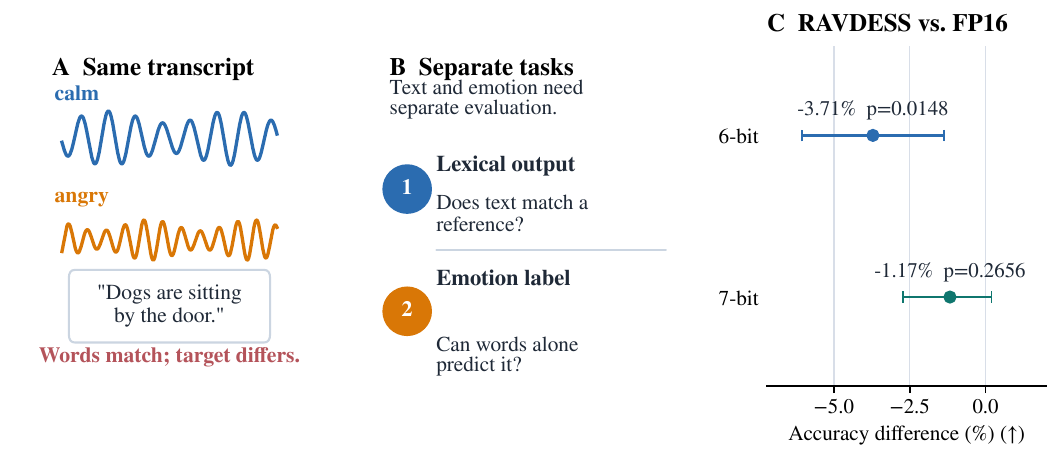}
\caption{Why quantized speech models need more than a transcript score. \textbf{A:} Schematic waveforms illustrate two recordings with the same words but different emotion labels. \textbf{B:} Translation scores do not measure performance on labels that the transcript cannot determine. \textbf{C:} The selected 6-bit allocation has lower RAVDESS accuracy than FP16, with an absolute difference of $-3.71$\% (95\% speaker interval $[-6.05,-1.37]$\%; exact two-sided sign-flip $p=0.0148$). The 7-bit interval includes zero.}
\label{fig:evaluation-protocol}
\end{figure}

We ask: \emph{what evidence is needed before text-output scores are used to describe quantized-model performance on speech tasks beyond the transcript?} We report translation scores, emotion recognition on a task whose labels are not recoverable from the transcript, a half-precision (FP16) baseline, and allocation controls with matched bit budgets. Figure~\ref{fig:evaluation-protocol} shows why these evaluations answer different questions.

\Needspace{6\baselineskip}
We apply this protocol to Qwen2-Audio \citep{qwen2audio2024}. Two mixed 4/8-bit allocations, averaging 6 and 7 bits per parameter, were selected using a disjoint speech-translation development split and fixed before test evaluation. We evaluate them on FLEURS and compare them with FP16, front-block, and stratified-uniform allocations on RAVDESS. The low-bit controls use the same average bit budgets.

\Needspace{5\baselineskip}
The selected 6-bit allocation has lower emotion accuracy than FP16, although its translation-score intervals include zero. The 7-bit difference from FP16 remains unresolved. These results illustrate why translation and emotion recognition require separate measurements.

\noindent\textbf{Contributions.} This paper makes three contributions.
\begin{enumerate}
  \item We state an evaluation protocol that separates text output from performance on a task whose labels cannot be recovered from the transcript. The protocol reports an FP16 baseline, matched bit budgets, and uncertainty at the independent-unit level.
  \item We evaluate the same fixed 6- and 7-bit Qwen2-Audio allocations on FLEURS and RAVDESS, including an FP16 reference and controls with matched bit budgets.
  \item We find no resolved FLEURS difference from FP16 for either selected allocation. On RAVDESS, the canonical-label 6-bit difference from FP16 is $-3.71$\% with a speaker interval excluding zero and exact sign-flip $p=0.0148$; the 7-bit interval includes zero. Neither selected allocation leads both matched controls.
\end{enumerate}

\section{Related Work}

\paragraph{Low-bit language and speech models.} LLM.int8(), GPTQ, SmoothQuant, and activation-aware weight quantization establish widely used PTQ approaches for large language models \citep{dettmers2022llmint8,frantar2022gptq,xiao2022smoothquant,lin2023awq}. Speech-specific work studies mixed-precision PTQ for automatic speech recognition, model compression for speech translation, selected-layer codec compression, and PTQ for speech language models \citep{genptq2025,ultralowptqspeech2025,efficientstcompression2025,selectedcodec2026,dietkit2026}. GenPTQ studies mixed-precision ASR allocation and lexical recognition, whereas we evaluate fixed Qwen2-Audio allocations across text output and a transcript-nondiagnostic label endpoint without proposing an allocator.

\paragraph{Evaluating audio use.} Audio-language evaluation distinguishes direct audio use from text-mediated answers and studies temporal grounding and token-level readout \citep{directmediated2026,tagbench2026,tempo2026,readouttokens2026,temporalfailures2026}. LISTEN intervenes on aligned and conflicting lexical and acoustic emotion cues across six audio-language models \citep{chen-etal-2026-audio}. Our balanced-sentence design instead holds empirical label frequencies equal across two sentences; it cannot isolate acoustic cues. We ask whether text-output evidence transfers to fixed, quantized Qwen2-Audio allocations evaluated against FP16 and same-budget controls, without identifying cue use or proposing a new audio-use benchmark.

\section{Evaluation Protocol}
\label{sec:eval-protocol}

\paragraph{Datasets and comparison units.} We evaluate translation on 508 English-to-German FLEURS utterances, using the same examples for FP16 and both selected allocations. Emotion recognition uses 512 RAVDESS clips from 16 speakers, with 32 clips per speaker. All seven configurations use the same clips: FP16 and the selected, front-block, and stratified-uniform allocations at 6 and 7 bits. Accuracy is measured over clips; uncertainty is estimated over speakers to account for repeated recordings from each actor.

\subsection{Two different questions}

\paragraph{Text-output quality.} Automatic speech recognition (ASR) and speech translation (ST) evaluate lexical output. This study reports BLEU and chrF for English-to-German speech translation. These scores describe the generated text; they do not evaluate an emotion label that the transcript cannot determine.

\paragraph{A task whose label the transcript cannot determine.} We report emotion recognition on a test set where the same two sentences occur with each of eight emotion labels. The words therefore do not determine the target, but this property does not establish that the model relies on acoustic cues; that would require transcript-only input or controlled audio perturbations. All low-bit alternatives use matched parameter-weighted bit budgets. We compare the translation-selected plan with front-block and stratified-uniform allocations that change which modules receive each bit width while holding the budget nearly fixed. The experiment applies dequantized per-channel round-to-nearest weights and does not measure a packed implementation or runtime efficiency.

\paragraph{What sentence balance establishes.} Within the evaluated RAVDESS subset, both spoken sentences appear equally often with each of the eight labels. A rule that receives only the sentence identity therefore sees the same empirical label distribution for either sentence, and its best sentence-only accuracy is 12.5\% on this balanced set. This is a property of the test design, not a measured transcript-only model baseline. Emotion accuracy above 12.5\% shows that the evaluated system uses information beyond the identity of these two sentences, but the available outputs cannot identify whether that information came from acoustic emotion cues, speaker or recording characteristics, or another route. Figure~\ref{fig:evaluation-protocol} depicts this distinction. We consequently describe the task as lexically non-diagnostic and describe the measured outcome as emotion-label accuracy; we do not call it a causal test of waveform reliance.

\ifdefined\ARXIVCOPY
\paragraph{A formal view of the task property.} Let $S$ be the identity of the spoken sentence and $Y$ the acted emotion label. The empirical distribution of this test set satisfies
\begin{equation}
\widehat{P}(Y=e\mid S=s)=\frac{1}{8}
\quad\text{for every evaluated sentence }s\text{ and label }e.
\label{eq:sentence-balance}
\end{equation}
A decision rule that predicts one of the eight labels using only $S$ has expected accuracy $1/8$ under this empirical distribution. Randomizing the prediction does not improve this expectation; allowing invalid outputs gives an upper bound of $1/8$. This calculation refers to the literal sentence identity and excludes speaker information, timing, or punctuation inferred from audio. Above-chance accuracy motivates an examination of non-lexical information, while identifying the information used requires an intervention. Sentence balance justifies the evaluation target, without assigning observed accuracy to a specific acoustic mechanism.
\fi

\paragraph{Three comparisons with different meanings.} The FP16 comparison measures the accuracy change for a specific quantized allocation under a fixed task and prompt. A same-budget comparison asks whether the placement of low and high precision matters beyond its average bit width; matching that average does not match each module's quantization error or guarantee equal activation behavior. The cross-task comparison asks whether a favorable translation result is sufficient to infer preserved performance on a task whose target is absent from the transcript. Its two tasks have different datasets, prompts, metrics, and output units, so their score differences cannot be subtracted or treated as a common effect size. The design supports a failure of inference from one endpoint to the other when the observed results disagree. It does not estimate the effect of changing the task while holding inputs or labels fixed.

\subsection{Qwen2-Audio case study}

\paragraph{Model and allocations.} We use Qwen2-Audio-7B-Instruct (revision \texttt{0a095220c30b}). For each budget, we selected the highest-chrF allocation among seven predefined candidates on a separate 387-example speech-translation development split and fixed it before either test evaluation. RAVDESS was not used to choose an allocation. Each plan starts modules at 4 bits, fixes the output head at 8 bits, and upgrades whole modules to 8 bits subject to its parameter-weighted budget.

\paragraph{Matched controls.} The front-block control assigns 8-bit precision to modules in model order. The stratified-uniform control instead assigns 8-bit precision by visiting block midpoints recursively across depth. Both use 4- and 8-bit modules, but assign high precision to different modules from the selected allocation. The selected 6- and 7-bit allocations average 5.996 and 6.996 bits, respectively; their matched controls average 5.998 and 6.994 bits. The largest gap is less than 0.0018 bits before rounding. These comparisons test the placement of precision at effectively fixed average precision.

\ifdefined\ARXIVCOPY
\paragraph{Parameter weighting and allocation selection.} For the modules covered by an allocation, let $n_m$ be the number of weight parameters in module $m$ and let $b_m\in\{4,8\}$ be its assigned precision. The reported average is
\begin{equation}
\bar b=\frac{\sum_m n_m b_m}{\sum_m n_m}.
\label{eq:weighted-budget}
\end{equation}
Weighting by parameter count matters because assigning 8 bits to one large matrix changes the budget more than assigning 8 bits to one small matrix. Both controls retain the 8-bit output head and skip an upgrade when its parameter-weighted cost exceeds the remaining budget. Modules within a block need not share the same precision. Equation~\ref{eq:weighted-budget} fixes the resource constraint for the comparison; it does not equate the reconstruction errors of different allocations.

The selected candidate maximizes development chrF within the seven-allocation pool. This makes it a translation-selected allocation, without establishing an optimum for emotion recognition or for allocations outside the pool. The case study examines how this fixed choice performs on the second task. The front-block and stratified-uniform controls provide alternatives whose ordering is defined structurally. Their role is to test whether the observed result depends on precision placement after accounting for the average budget.

\paragraph{Weight reconstruction.} Round-to-nearest quantization approximates weights at each module's assigned precision, then reconstructs floating-point weights for inference. It changes their numerical values without using a packed integer implementation. The same average $\bar b$ therefore permits different weight approximations; the matched controls measure how those placements affect task accuracy.
\fi

\paragraph{Translation task and scoring.} We evaluate English-to-German speech translation on 508 held-out FLEURS utterances \citep{conneau2022fleurs,seamlessm4t2023}. BLEU is the primary metric and chrF is a complementary character-level metric, both computed with SacreBLEU 2.5.1 using one reference translation per utterance \citep{post2018call,popovic2015chrf}. BLEU preserves case and uses the 13a tokenizer, exponential smoothing, and a fixed maximum order of four. chrF also preserves case and uses character n-grams up to order six, with no word n-grams and with whitespace ignored. Its effective-order setting averages only over orders present in both the prediction and reference.

\paragraph{RAVDESS emotion evaluation.} RAVDESS contains acted speech recordings with eight emotion labels \citep{livingstone2018ravdess}. We evaluate 512 clips from actors 1--16, with four clips for every actor--emotion pair. Each of the two spoken sentences occurs 32 times for every emotion, so sentence identity is non-diagnostic within this set. The prompt asks the model to choose exactly one of neutral, calm, happy, sad, angry, fearful, disgust, or surprised. Canonical scoring lowercases output, removes periods and commas, and maps the unambiguous morphological form ``disgusted'' to ``disgust'' before requiring exactly one listed label. All arms use greedy decoding, deterministic operations, seed 1729, and an eight-token limit. Sentence balance alone does not distinguish emotion cues from speaker or recording cues.

\ifdefined\ARXIVCOPY
\paragraph{Decoding and the outcome definition.} Scoring extracts emotion-label tokens from the normalized output and accepts exactly one, ignoring other words. For example, ``The emotion is happy'' is scored as happy; ``happy or sad'' is incorrect. A response with no recognized label is also incorrect. The same conversion is applied to every configuration. Accepting an unambiguous inflection avoids counting that surface variation as an emotion error. The outcome evaluates categorical label agreement, without scoring an explanation, confidence, or listener judgment. Greedy decoding holds the decoding rule fixed. The intervals do not estimate variation across prompts, label vocabularies, or decoding policies.
\fi

FLEURS paired intervals use 2,000 utterance bootstrap resamples with seed 1729. RAVDESS intervals use an equal-weight bootstrap over 16 speakers with 10,000 resamples. We additionally enumerate all $2^{16}$ sign assignments of the paired speaker means for a two-sided exact sign-flip test. This test is valid under the sharp null that a paired speaker-level difference is sign-exchangeable. We report its $p$ values as a complement to the descriptive speaker bootstrap, not as a substitute for the interval or as a multiplicity-adjusted result.

\paragraph{Paired estimands and analysis units.} Every reported difference has the order selected allocation minus a named comparator. For FLEURS, each bootstrap draw selects utterance identities with replacement and retains the paired prediction--reference records before recomputing BLEU and chrF. These intervals describe variation over the 508 stored outputs, conditional on this scoring procedure; they do not account for a new generation run. For RAVDESS, each draw selects the 16 speakers with replacement and retains all clips for a selected speaker. The resulting interval avoids treating repeated recordings from an actor as independent. Both intervals use the 2.5th and 97.5th percentiles. An interval including zero leaves that named comparison unresolved; it does not establish equal performance.

\ifdefined\ARXIVCOPY
\paragraph{Emotion accuracy as a paired speaker mean.} Write $\hat y_{g,s,i}$ for the normalized prediction from configuration $g$ on clip $i$ of speaker $s$, and $y_{s,i}$ for its reference label. The speaker's accuracy is
\begin{equation}
a_{g,s}=\frac{1}{32}\sum_{i=1}^{32}
\mathbf{1}\{\hat y_{g,s,i}=y_{s,i}\}.
\label{eq:speaker-accuracy}
\end{equation}
For a selected allocation $g$ and comparator $r$, the absolute accuracy difference in percent is
\begin{equation}
\Delta_{g,r}=100\times\frac{1}{16}\sum_{s=1}^{16}
\bigl(a_{g,s}-a_{r,s}\bigr).
\label{eq:paired-accuracy}
\end{equation}
Every speaker contributes the same number of clips, so the equally weighted speaker mean also equals overall clip accuracy. Resampling individual clips would omit dependence among recordings from one actor. Pairing retains the same speakers and clips on both sides of Equation~\ref{eq:paired-accuracy}. In contrast, BLEU and chrF are recomputed as corpus scores in each paired utterance resample; their intervals are not averages of sentence-level score differences.
\fi

\paragraph{Multiple comparisons.} We report two selected-versus-FP16 comparisons and four matched-control comparisons on RAVDESS. The intervals and sign-flip tests are not adjusted for multiple comparisons. Each describes its named pair of configurations, so a favorable result for one pair does not establish superiority across all tested allocations.

\section{Results}

\subsection{Translation-score differences from FP16 remain unresolved}

Table~\ref{tab:cross-endpoint-results} reports FLEURS translation scores and RAVDESS emotion accuracy for the same selected allocations. Each task uses its own prompt, target, and metric; the scores are not on a common scale.

\begin{table}[h]
\centering
\caption{FLEURS translation scores and RAVDESS emotion accuracy for the same precision allocations. BLEU and chrF use SacreBLEU 2.5.1; emotion accuracy (\%) uses the label-normalization rule in Section~\ref{sec:eval-protocol}. Differences are calculated before rounding. Shading identifies the translation-selected allocations.}
\label{tab:cross-endpoint-results}
\vspace{6pt}
\small
\setlength{\tabcolsep}{3pt}
\begin{tabularx}{0.92\linewidth}{@{}L{0.32\linewidth} C{0.13\linewidth} C{0.13\linewidth} C{0.19\linewidth}@{}}
\toprule
\textbf{Allocation} & \textbf{BLEU $\uparrow$} & \textbf{chrF $\uparrow$} & \textbf{Accuracy (\%) $\uparrow$} \\
\midrule
FP16 & 20.36 & 49.73 & 70.12 \\
\rowcolor{selectedrow} Translation-selected, 6 bit & 20.22 & 49.60 & 66.41 \\
\rowcolor{selectedrow} Translation-selected, 7 bit & 20.33 & 49.68 & 68.95 \\
\bottomrule
\end{tabularx}
\end{table}

\begin{figure}[h]
\centering
\includegraphics[width=\linewidth]{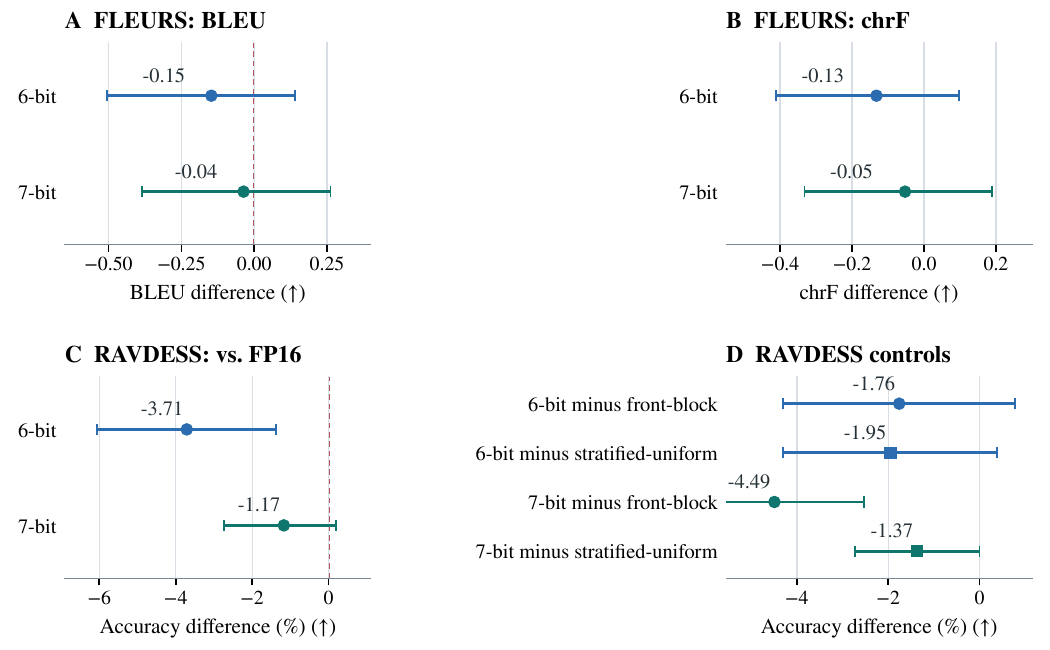}
\caption{Selected allocation minus comparator with 95\% intervals. Panels A and B show FLEURS translation-score differences from FP16, using paired utterance resampling (508 examples; 2,000 resamples). Panels C and D show RAVDESS accuracy differences from FP16 and same-budget controls, using speaker resampling (16 speakers; 10,000 resamples). A negative value favors the comparator.}
\label{fig:prompt-matched-bridge}
\end{figure}

For 6 bits, selected minus FP16 is $-0.15$ BLEU (paired 95\% interval $[-0.51,+0.14]$) and $-0.13$ chrF ($[-0.41,+0.10]$). For 7 bits, it is $-0.04$ BLEU ($[-0.38,+0.26]$) and $-0.05$ chrF ($[-0.33,+0.19]$). Every FLEURS interval includes zero. Figure~\ref{fig:prompt-matched-bridge} and Table~\ref{tab:deterministic-fleurs} show these results alongside the endpoint-specific RAVDESS comparisons. The intervals do not show that the outputs are interchangeable with FP16; they leave the measured differences unresolved on this fixed test list.

\ifdefined\ARXIVCOPY
\paragraph{Unresolved differences and preservation.} The four translation intervals contain both negative and positive differences. They are therefore compatible with a small loss or gain under the stated resampling procedure. Establishing preservation to a practical tolerance would require an acceptable loss margin specified for each metric before examining these comparisons. A non-inferiority interpretation would then compare the lower interval bound with that margin; an equivalence interpretation would require a corresponding upper bound as well. This study specifies no such margins. Its FLEURS result is an unresolved difference, and the emotion evaluation remains necessary even if a future translation analysis meets an explicit preservation criterion. Translation adequacy and agreement with an emotion label are separate requirements for a speech model used in both settings.
\fi

\begin{table}[h]
\centering
\caption{FLEURS translation-score differences, selected allocation minus FP16. Paired 95\% bootstrap intervals resample the 508 utterances 2,000 times with seed 1729. All intervals include zero. CI denotes confidence interval.}
\label{tab:deterministic-fleurs}
\vspace{6pt}
\small
\setlength{\tabcolsep}{4pt}
\begin{tabularx}{0.82\linewidth}{@{}L{0.25\linewidth} Y Y@{}}
\toprule
\textbf{Allocation} & \textbf{$\Delta$BLEU $\uparrow$ (95\% CI)} & \textbf{$\Delta$chrF $\uparrow$ (95\% CI)} \\
\midrule
6 bit & $-0.15$ $[-0.51,+0.14]$ & $-0.13$ $[-0.41,+0.10]$ \\
7 bit & $-0.04$ $[-0.38,+0.26]$ & $-0.05$ $[-0.33,+0.19]$ \\
\bottomrule
\end{tabularx}
\end{table}

\Needspace{12\baselineskip}
\subsection{RAVDESS identifies a 6-bit FP16 deficit}

The RAVDESS columns in Table~\ref{tab:cross-endpoint-results} use the same selected allocations as the translation evaluation. Figure~\ref{fig:prompt-matched-bridge} shows their comparisons with FP16 and matched controls. At 6 bits, the selected allocation reaches 66.41\% accuracy. Its absolute difference from FP16 is $-3.71$\% (95\% speaker interval $[-6.05,-1.37]$\%; two-sided exact sign-flip $p=0.0148$). At 7 bits, accuracy is 68.95\%, with an absolute difference of $-1.17$\% from FP16 ($[-2.73,+0.20]$\%; $p=0.2656$). The 6-bit allocation therefore has lower emotion accuracy under this setup, while the 7-bit difference remains unresolved.

\ifdefined\ARXIVCOPY
\Needspace{10\baselineskip}
\paragraph{Reading the accuracy change.} FP16 predicts 359 of the 512 labels correctly, compared with 340 for the selected 6-bit allocation and 353 for the selected 7-bit allocation. The two deficits are therefore net differences of 19 and six correct predictions on the same clips. A net difference does not describe every changed decision: errors corrected by an allocation and errors introduced by it may offset one another. Table~\ref{tab:per-speaker-results} supplies the corresponding speaker-level view. The 6-bit decrease appears across 12 of the 16 actors, while nine actors have no net accuracy change at 7 bits. Neither count locates the responsible modules or identifies the type of emotion cue affected. It describes the distribution of the observed accuracy change across the evaluated speakers.
\fi

\Needspace{13\baselineskip}
\paragraph{What the controls show.} At 6 bits, the selected-minus-front-block absolute accuracy difference is $-1.76$\% ($[-4.30,+0.78]$\%; $p=0.2754$) and the selected-minus-stratified-uniform difference is $-1.95$\% ($[-4.30,+0.39]$\%; $p=0.1836$). At 7 bits, the selected-minus-front-block difference is $-4.49$\% ($[-6.45,-2.54]$\%; $p=0.0020$) and the selected-minus-stratified-uniform difference is $-1.37$\% ($[-2.73,0.00]$\%; $p=0.1465$). The selected plan has a lower point estimate than both controls at both budgets. Only the 7-bit front-block comparison has an interval that excludes zero. Neither selected allocation has the highest observed accuracy at its bit budget, and several control differences remain unresolved. Appendix~\ref{app:ravdess-comparisons} reports every allocation's accuracy and its corresponding contrasts.

\subsection{Speaker-level decomposition}

Table~\ref{tab:per-speaker-results} shows how the accuracy differences vary across speakers. The 6-bit difference is negative for 12 actors, zero for two, and positive for two. The 7-bit difference is negative for five actors, zero for nine, and positive for two. These actor-level means are the units used in the speaker bootstrap and sign-flip analysis.

\ifdefined\ARXIVCOPY
\Needspace{9\baselineskip}
\fi

The exact sign-flip calculation uses one paired mean difference per actor. Under its sharp null, each actor-level difference may be multiplied by either $+1$ or $-1$ with equal probability. Enumerating the $65{,}536$ possible signs gives $p=0.0148$ for 6-bit selected minus FP16 and $p=0.2656$ for 7-bit selected minus FP16. This calculation does not assume that clips are independent, but it does rely on sign-exchangeability of the actor-level differences under the null. The speaker bootstrap and exact test agree on the qualitative reading of the two FP16 contrasts: a resolved 6-bit deficit and an unresolved 7-bit difference.

\begin{table}[t]
\centering
\caption{RAVDESS emotion accuracy by actor. Each actor contributes 32 clips. The final two columns are absolute accuracy differences, selected allocation minus FP16, for that actor, computed before rounding. The uncertainty analyses use these actor-level means.}
\label{tab:per-speaker-results}
\vspace{6pt}
\small
\setlength{\tabcolsep}{3pt}
\begin{tabularx}{0.96\linewidth}{@{}C{0.10\linewidth} C{0.15\linewidth} C{0.15\linewidth} C{0.15\linewidth} C{0.15\linewidth} C{0.15\linewidth}@{}}
\toprule
\textbf{Actor} & \textbf{FP16 \mbox{(\%) $\uparrow$}} & \textbf{Selected 6 \mbox{(\%) $\uparrow$}} & \textbf{Selected 7 \mbox{(\%) $\uparrow$}} & \textbf{6 $-$ FP16 \mbox{(\%) $\uparrow$}} & \textbf{7 $-$ FP16 \mbox{(\%) $\uparrow$}} \\
\midrule
01 & 56.25 & 53.12 & 53.12 & $-3.12$ & $-3.12$ \\
02 & 75.00 & 65.62 & 75.00 & $-9.38$ & 0.00 \\
03 & 84.38 & 78.12 & 84.38 & $-6.25$ & 0.00 \\
04 & 68.75 & 62.50 & 68.75 & $-6.25$ & 0.00 \\
05 & 71.88 & 68.75 & 71.88 & $-3.12$ & 0.00 \\
06 & 65.62 & 62.50 & 65.62 & $-3.12$ & 0.00 \\
07 & 81.25 & 78.12 & 84.38 & $-3.12$ & $+3.12$ \\
08 & 68.75 & 56.25 & 62.50 & $-12.50$ & $-6.25$ \\
09 & 65.62 & 65.62 & 62.50 & 0.00 & $-3.12$ \\
10 & 59.38 & 62.50 & 62.50 & $+3.12$ & $+3.12$ \\
11 & 75.00 & 81.25 & 75.00 & $+6.25$ & 0.00 \\
12 & 71.88 & 59.38 & 62.50 & $-12.50$ & $-9.38$ \\
13 & 65.62 & 62.50 & 62.50 & $-3.12$ & $-3.12$ \\
14 & 68.75 & 68.75 & 68.75 & 0.00 & 0.00 \\
15 & 68.75 & 65.62 & 68.75 & $-3.12$ & 0.00 \\
16 & 75.00 & 71.88 & 75.00 & $-3.12$ & 0.00 \\
\bottomrule
\end{tabularx}
\end{table}

\ifdefined\ARXIVCOPY
\Needspace{10\baselineskip}
\fi

\section{Discussion}

The FLEURS and RAVDESS evaluations answer different empirical questions. In FLEURS, the selected-minus-FP16 intervals include zero for both BLEU and chrF at both budgets. In RAVDESS, the selected 6-bit allocation has lower accuracy than FP16 under both uncertainty analyses; the 7-bit difference is unresolved. Translation evaluation therefore does not establish whether emotion accuracy is preserved.

The matched controls sharpen that interpretation. FP16 is a useful reference for the loss incurred by one allocation, but it does not show whether the selected placement is distinguished from other placements at the same average bit width. The control outcomes identify lower selected-plan point estimates at both budgets and a resolved adverse 7-bit front-block comparison. They leave the remaining control pairs unresolved. This combination is more informative than a single reported text score, while remaining conditional on one checkpoint, two selected budgets, a fixed prompt, and 16 speakers.

\ifdefined\ARXIVCOPY
\paragraph{What the cross-task comparison supports.} We test the inference that small text-score changes establish preserved behavior on another speech task. Evaluation of a target beyond the words supplies information absent from those scores. Using the same allocation connects both outcomes to a concrete model configuration. The tasks have different populations and outputs, so their comparison does not isolate a shared acoustic representation or establish a causal translation--emotion trade-off.

The controls also caution against treating the selected allocation as a general prescription for precision placement. Each allocation changes the reconstructed weights in multiple modules at once. Matching $\bar b$ removes average precision as a substantial difference between the low-bit comparisons, but it does not identify a module-level cause. A causal localization study would require interventions on specified modules with the remaining allocation held fixed. The present comparison is sufficient to show that the translation-selected placement has no observed emotion-accuracy advantage over both tested controls. It leaves the design of an allocation that serves both tasks as a separate question.
\fi

\Needspace{7\baselineskip}
RAVDESS sentence balance makes the endpoint lexically non-diagnostic, but it does not prove that the model uses emotion-related acoustic cues. Speaker identity, recording conditions, or other non-lexical information may also contribute. The study applies dequantized weights and has no packed-model measurement, so nominal bit width does not establish memory savings, latency, or throughput. These boundaries motivate task-specific evaluation; they do not describe deployment behavior.

\paragraph{Evaluating an intended speech task.} If the intended use involves an emotion label or another target absent from the transcript, evaluate that target with an appropriate full-precision baseline and controls at the same bit budget. A translation score cannot replace this measurement. The present result does not establish a general trade-off between translation and emotion recognition, or predict the behavior of other quantized speech models.

\paragraph{Further tests of cue use and scope.} A transcript-only condition would measure whether information outside the words improves emotion-label predictions under the same setup. Audio perturbations that preserve spoken words while altering candidate cues could test which information the model uses. Evaluation across checkpoints and naturalistic emotion data would test the scope of the accuracy result. Packed implementations would require separate measurements of memory use, latency, and throughput.

\ifdefined\ARXIVCOPY
For an application requiring both translation and emotion output, allocation selection also needs an explicit objective for each task. Those objectives have different units, so combining them requires a stated weighting or an acceptable loss for each outcome. Such choices belong in development evaluation, followed by separate test comparisons. The present study uses translation chrF for selection and evaluates emotion accuracy afterward. It reveals a limitation of that selection objective for the tested allocations, without specifying a preferred weighting or establishing the best allocation for a joint objective.
\fi

\section{Limitations}

The experiments use one Qwen2-Audio checkpoint and one acted North American English emotion set. Results may differ for spontaneous emotion, long-form speech, multilingual prosody, or other tasks whose labels cannot be recovered from the transcript. The FLEURS intervals describe variation over test utterances for one generation run; they do not estimate variability across repeated generation runs. The RAVDESS analysis has 16 speaker clusters, and its exact sign-flip test assumes sign-exchangeability under the null. Sentence balance does not distinguish emotion cues from speaker or recording cues, and it does not establish causal audio use. The experiments do not measure memory use, latency, or throughput with packed low-bit weights.

No human listening or human emotion judgment was performed for this study. Exact-match accuracy measures agreement with the dataset's acted emotion labels under the specified prompt; it does not establish perceived emotion quality or usefulness to listeners.

\section{Conclusion}

Text-output scores alone do not establish quantized-model performance on a task whose labels cannot be recovered from the transcript. In this Qwen2-Audio case study, the translation-score differences from FP16 remain unresolved at 6 and 7 bits. On RAVDESS, the selected 6-bit allocation has lower accuracy than FP16, while the 7-bit interval includes zero. Neither selected allocation has higher observed accuracy than both same-budget controls. These results support separate evaluation of text quality and the intended speech task, using a full-precision reference, matched budgets, and uncertainty estimates that account for speaker dependence.

Future work includes evaluation across checkpoints and naturalistic emotion data, transcript-only conditions and controlled audio perturbations to study cue use, and human evaluation of perceived emotion understanding.

\ifdefined\ARXIVCOPY
\clearpage
\fi

\section*{AI Use Statement}
We used generative AI tools to help find related work, draft and edit text, review the evaluation and its interpretation, and revise figure code and layout. We checked citations, numerical results, and the rendered paper against the cited sources and stored experiment outputs. The authors take responsibility for the final manuscript and artifacts.

\section*{Reproducibility Statement}
Section~\ref{sec:eval-protocol} describes the test populations, precision allocations, scoring rules, and uncertainty analyses. The reproducibility repository linked in the introduction provides the allocation definitions, scoring code, and metric configurations.

\setlength{\bibsep}{0pt plus 0.2ex}
\bibliographystyle{iclr2027_conference}
\bibliography{references}

@inproceedings{dietkit2026,
  author = {Liu, Danni and Koneru, Sai and Niehues, Jan},
  title = {{Diet-KIT}: Post-Training Quantization for Speech {LLM}s},
  booktitle = {Proceedings of the 23rd International Conference on Spoken Language Translation},
  year = {2026},
  pages = {189--196},
  doi = {10.18653/v1/2026.iwslt-1.21},
  url = {https://aclanthology.org/2026.iwslt-1.21/}
}

@inproceedings{selectedcodec2026,
  author = {Palomino, Alonso},
  title = {Selected-Layer Codec Compression for Compact Speech Translation Models: An {IWSLT} 2026 English-to-Chinese Submission},
  booktitle = {Proceedings of the 23rd International Conference on Spoken Language Translation},
  year = {2026},
  pages = {40--46},
  doi = {10.18653/v1/2026.iwslt-1.4},
  url = {https://aclanthology.org/2026.iwslt-1.4/}
}

@misc{efficientstcompression2025,
  author = {Moslem, Yasmin},
  title = {Efficient Speech Translation through Model Compression and Knowledge Distillation},
  year = {2025},
  eprint = {2505.20237},
  archivePrefix = {arXiv},
  url = {https://arxiv.org/abs/2505.20237}
}

@misc{qwen2audio2024,
  author = {Chu, Yunfei and Xu, Jin and Yang, Qian and Wei, Haojie and Wei, Xipin and Guo, Zhifang and Leng, Yichong and Lv, Yuanjun and He, Jinzheng and Lin, Junyang and Zhou, Chang and Zhou, Jingren},
  title = {{Qwen2-Audio} Technical Report},
  year = {2024},
  eprint = {2407.10759},
  archivePrefix = {arXiv},
  url = {https://arxiv.org/abs/2407.10759}
}

@inproceedings{conneau2022fleurs,
  author = {Conneau, Alexis and Ma, Min and Khanuja, Simran and Zhang, Yu and Axelrod, Vera and Dalmia, Siddharth and Riesa, Jason and Rivera, Clara and Bapna, Ankur},
  title = {{FLEURS}: Few-shot Learning Evaluation of Universal Representations of Speech},
  booktitle = {2022 IEEE Spoken Language Technology Workshop},
  year = {2022},
  pages = {798--805},
  doi = {10.1109/SLT54892.2023.10023141}
}

@misc{seamlessm4t2023,
  author = {{Seamless Communication} and Barrault, Loic and Chung, Yu-An and Meglioli, Mariano Cora and Dale, David and Dong, Ning and Duquenne, Paul-Ambroise and Elsahar, Hady and others},
  title = {{SeamlessM4T}: Massively Multilingual and Multimodal Machine Translation},
  year = {2023},
  eprint = {2308.11596},
  archivePrefix = {arXiv},
  primaryClass = {cs.CL},
  url = {https://arxiv.org/abs/2308.11596}
}

@inproceedings{post2018call,
  author = {Post, Matt},
  title = {A Call for Clarity in Reporting {BLEU} Scores},
  booktitle = {Proceedings of the Third Conference on Machine Translation: Research Papers},
  year = {2018},
  pages = {186--191},
  address = {Brussels, Belgium},
  publisher = {Association for Computational Linguistics},
  doi = {10.18653/v1/W18-6319},
  url = {https://aclanthology.org/W18-6319/}
}

@inproceedings{popovic2015chrf,
  author = {Popovi\v{c}, Maja},
  title = {{chrF}: Character n-gram {F}-score for Automatic {MT} Evaluation},
  booktitle = {Proceedings of the Tenth Workshop on Statistical Machine Translation},
  year = {2015},
  pages = {392--395},
  address = {Lisbon, Portugal},
  publisher = {Association for Computational Linguistics},
  doi = {10.18653/v1/W15-3049},
  url = {https://aclanthology.org/W15-3049/}
}

@article{livingstone2018ravdess,
  author = {Livingstone, Steven R. and Russo, Frank A.},
  title = {The {Ryerson} Audio-Visual Database of Emotional Speech and Song ({RAVDESS}): A Dynamic, Multimodal Set of Facial and Vocal Expressions in North American English},
  journal = {PLOS ONE},
  year = {2018},
  volume = {13},
  number = {5},
  pages = {e0196391},
  doi = {10.1371/journal.pone.0196391}
}

@inproceedings{genptq2025,
  author = {Kang, Beom Jin and Kim, Hyun},
  title = {{GenPTQ}: Green Post-Training Quantization for Large-Scale {ASR} Models with Mixed-Precision Bit Allocation},
  booktitle = {Findings of the Association for Computational Linguistics: EMNLP 2025},
  year = {2025},
  pages = {10704--10718},
  publisher = {Association for Computational Linguistics},
  doi = {10.18653/v1/2025.findings-emnlp.566},
  url = {https://aclanthology.org/2025.findings-emnlp.566/}
}

@misc{tagbench2026,
  title = {{TAG-Bench}: Benchmarking Temporal Audio Grounding in Large Audio Language Models},
  author = {Dai, Yuhang and Shu, Xin and Li, Zengxi and Xie, Lei and Li, Xiangang and Yu, Jianwei},
  year = {2026},
  eprint = {2609.01542},
  archivePrefix = {arXiv},
  primaryClass = {cs.SD},
  url = {https://arxiv.org/abs/2609.01542}
}

@inproceedings{tempo2026,
  title = {{TEMPO}: Temporally-grounded Multi-task Post-training for Large Audio-Language Models},
  author = {Kulkarni, Apoorva and Jayakumar, Kaousheik and Ghosh, Sreyan and Aich, Utathya and Duraiswami, Ramani and Manocha, Dinesh},
  booktitle = {Proceedings of the 2026 Conference on Empirical Methods in Natural Language Processing},
  year = {2026},
  url = {https://openreview.net/forum?id=LoXjHBlPEd}
}

@misc{readouttokens2026,
  title = {From Semantics to Readout: Mechanistic Understanding of Audio Tokens after Fine-Tuning for Temporal Audio Grounding},
  author = {Ma, Yujian and Sang, Jinqiu and Li, Ruizhe and Yu, Jiaao and Li, Ang},
  year = {2026},
  url = {https://arxiv.org/abs/2607.25355}
}

@misc{temporalfailures2026,
  title = {A Closer Look at Failure Modes in Temporal Understanding of Large Audio-Language Models},
  author = {Kulkarni, Apoorva and Jayakumar, Kaousheik and Ghosh, Sreyan and Wiegreffe, Sarah and Duraiswami, Ramani and Manocha, Dinesh},
  year = {2026},
  url = {https://arxiv.org/abs/2606.17417}
}

@misc{directmediated2026,
  title = {Direct or Mediated? Task-Dependent Audio Information Routing in Large Audio Language Models},
  author = {Zhang, Yizhou and Zhou, Wangjin and Gu, Xin and Wang, Yichi and Tan, Wei and Zhao, Yi and Gong, Zhi and Imoto, Keisuke and Kawahara, Tatsuya},
  year = {2026},
  eprint = {2608.27026},
  archivePrefix = {arXiv},
  primaryClass = {cs.SD},
  url = {https://arxiv.org/abs/2608.27026}
}

@misc{dettmers2022llmint8,
  author = {Dettmers, Tim and Lewis, Mike and Belkada, Younes and Zettlemoyer, Luke},
  title = {{LLM.int8()}: 8-bit Matrix Multiplication for Transformers at Scale},
  year = {2022},
  eprint = {2208.07339},
  archivePrefix = {arXiv},
  url = {https://arxiv.org/abs/2208.07339}
}

@misc{frantar2022gptq,
  author = {Frantar, Elias and Ashkboos, Saleh and Hoefler, Torsten and Alistarh, Dan},
  title = {{GPTQ}: Accurate Post-Training Quantization for Generative Pre-trained Transformers},
  year = {2022},
  eprint = {2210.17323},
  archivePrefix = {arXiv},
  url = {https://arxiv.org/abs/2210.17323}
}

@misc{xiao2022smoothquant,
  author = {Xiao, Guangxuan and Lin, Ji and Seznec, Mickael and Wu, Hao and Demouth, Julien and Han, Song},
  title = {{SmoothQuant}: Accurate and Efficient Post-Training Quantization for Large Language Models},
  year = {2022},
  eprint = {2211.10438},
  archivePrefix = {arXiv},
  url = {https://arxiv.org/abs/2211.10438}
}

@misc{lin2023awq,
  author = {Lin, Ji and Tang, Jiaming and Tang, Haotian and Yang, Shang and Chen, Wei-Ming and Wang, Wei-Chen and Xiao, Guangxuan and Dang, Xingyu and Gan, Chuang and Han, Song},
  title = {{AWQ}: Activation-aware Weight Quantization for {LLM} Compression and Acceleration},
  year = {2023},
  eprint = {2306.00978},
  archivePrefix = {arXiv},
  url = {https://arxiv.org/abs/2306.00978}
}

@inproceedings{ultralowptqspeech2025,
  author = {Gu, Tianteng and Liu, Bei and Wang, Haoyu and Qian, Yanmin},
  title = {Ultra-Low Bit Post-Training Quantization of Large Speech Models via K-Means Clustering and Mixed Precision Allocation},
  booktitle = {Interspeech 2025},
  year = {2025},
  url = {https://www.isca-archive.org/interspeech_2025/gu25b_interspeech.html}
}

@inproceedings{chen-etal-2026-audio,
  title = {Do Audio {LLM}s Really {LISTEN}, or Just Transcribe? Measuring Lexical vs. Acoustic Emotion Cues Reliance},
  author = {Chen, Jingyi and Guo, Zhimeng and Chun, Jiyun and Wang, Pichao and Perrault, Andrew and Elsner, Micha},
  booktitle = {Proceedings of the 19th Conference of the European Chapter of the Association for Computational Linguistics},
  year = {2026},
  pages = {5848--5877},
  publisher = {Association for Computational Linguistics},
  doi = {10.18653/v1/2026.eacl-long.274},
  url = {https://aclanthology.org/2026.eacl-long.274/}
}

\clearpage
\appendix
\section{RAVDESS Allocation Comparisons}
\label{app:ravdess-comparisons}

Table~\ref{tab:ravdess-complete-arms} compares the selected allocations with FP16 and both controls at each bit budget. The selected allocations have lower observed accuracy than both controls. The intervals exclude zero for the 6-bit comparison with FP16 and the 7-bit comparison with front-block; the remaining differences are unresolved.

\begin{table}[H]
\centering
\caption{RAVDESS emotion accuracy and absolute accuracy differences, selected allocation minus comparator. Each interval is a 95\% speaker bootstrap interval over 16 actors with 10,000 resamples. The two-sided exact sign-flip test uses the paired actor means; its $p$ values are not adjusted for multiple comparisons. Differences are calculated before rounding. Shading identifies the selected allocations.}
\label{tab:ravdess-complete-arms}
\vspace{6pt}
\small
\setlength{\tabcolsep}{3pt}
\begin{tabularx}{0.96\linewidth}{@{}L{0.27\linewidth} C{0.14\linewidth} C{0.31\linewidth} C{0.12\linewidth}@{}}
\toprule
\textbf{Allocation} & \textbf{Accuracy \mbox{(\%) $\uparrow$}} & \textbf{Selected $-$ comparator \mbox{(\%) $\uparrow$} (95\% CI)} & \textbf{Sign-flip $p$} \\
\midrule
\multicolumn{4}{@{}l}{\textbf{6-bit budget}} \\
\addlinespace[3pt]
FP16 & 70.12 & $-3.71$ $[-6.05,-1.37]$ & 0.0148 \\
\rowcolor{selectedrow} Selected & 66.41 & Reference & --- \\
Front-block & 68.16 & $-1.76$ $[-4.30,+0.78]$ & 0.2754 \\
Stratified-uniform & 68.36 & $-1.95$ $[-4.30,+0.39]$ & 0.1836 \\
\addlinespace[7pt]
\multicolumn{4}{@{}l}{\textbf{7-bit budget}} \\
\addlinespace[3pt]
FP16 & 70.12 & $-1.17$ $[-2.73,+0.20]$ & 0.2656 \\
\rowcolor{selectedrow} Selected & 68.95 & Reference & --- \\
Front-block & 73.44 & $-4.49$ $[-6.45,-2.54]$ & 0.0020 \\
Stratified-uniform & 70.31 & $-1.37$ $[-2.73,0.00]$ & 0.1465 \\
\bottomrule
\end{tabularx}
\end{table}

The front-block allocation has the highest observed accuracy at 7 bits, while stratified-uniform has the highest at 6 bits. These rankings describe the observed scores. The unresolved intervals do not establish that either control is better than the selected allocation in those comparisons, and the unadjusted tests do not provide a familywise guarantee.

\end{document}